\documentclass[useAMS,usenatbib]{mn2e}

\usepackage{graphicx}                   
\usepackage{float}
\input{epsf.sty}                        
\input{psfig.sty}
\begin{document}

\title[Lennard-Jones quark matter and massive quark stars]
{Lennard-Jones quark matter and massive quark stars}

\author[Lai, Xu]{X. Y. Lai and R. X. Xu\\
School of Physics and State Key Laboratory of Nuclear Physics and
Technology, Peking University, Beijing 100871, China}

\maketitle

\begin{abstract}
Quark clustering could occur in cold quark matter because of the
strong coupling between quarks at realistic baryon densities of
compact stars.
Although one may still not be able to calculate this conjectured
matter from first principles, the inter-cluster interaction might be
analogized to the interaction between inert molecules. Cold quark
matter would then crystallize in a solid state if the inter-cluster
potential is deep enough to trap the clusters in the wells.
We apply the Lennard-Jones potential to describe the inter-cluster
potential, and derive the equations of state, which are stiffer than
that derived in conventional models (e.g., MIT bag model).
If quark stars are composed of Lennard-Jones matter, they could have
high maximum masses ($>2M_{\odot}$) as well as very low masses
($<10^{-3}M_{\odot}$). These features could be tested by
observations.
\end{abstract}


\section{Introduction}

To understand the nature of pulsars we need to know the state of
cold quark matter, in which the dominant degree of freedom is
quarks, and their Fermi energy is much larger than their thermal
energy.
However, this is a difficult task because of (i) the
non-perturbative effect of the strong interaction between quarks at
low energy scale and (ii) the many-body problem of vast assemblies
of interacting particles.

On one hand, some efforts have been made for understanding the
behavior of quantum chromo-dynamics (QCD) at high density, among
which a color super-conductivity (CSC) state is currently focused on
in perturbative QCD as well as in QCD-based effective
models~\citep[e.g.,][]{csc08}.
On the other hand, it is phenomenologically conjectured that
astrophysical cold quark matter could be in a solid
state~\citep{xu03}, since the strong interaction may render quarks
grouped in clusters and the ground state of realistic quark matter
might not be that of Fermi gas~\citep[see a recent discussion given
by][]{xu08}.
If the residua interaction between quark clusters is stronger than
their kinetic energy, each quark cluster could be trapped in the
potential well and cold quark matter will be in a solid state.
Solid quark stars still cannot be ruled out in both astrophysics and
particle physics~\citep{Horvath05,Owen05}.
Additionally, there is evidence that the interaction between quarks
is very strong in hot quark-gluon plasma~\citep[i.e., the strongly
coupled quark-gluon plasma,][]{Shuryak}, according to the recent
achievements of relativistic heavy ion collision experiments.
When the temperature goes down, it is reasonable to conjecture that
the interaction between quarks should be stronger than that in the
hot quark-gluon plasma.

Because of the difficulty to obtain a realistic state equation of
cold quark matter at a few nuclear densities, we try to apply some
phenomenological models, which would have some implications about
the properties of QCD at low energy scale if the astronomical
observations can provide us with some limitations on such models.
In our previous paper~\citep{lx09}, a polytropic quark star model
has been suggested in order to establish a general framework in
which theoretical quark star models could be tested by observations.
This model can help us to understand the observed masses of pulsars
and the energy released during some extreme bursts; however, this is
a phenomenological model and does not include the form of
interaction between quarks.
To calculate the interaction between quarks and predict the state of
matter for quark stars by QCD calculations is a difficult task;
however, it is still meaningful for us to consider some models to
explore the properties of quarks at the low energy scale.

We can compare the interaction between quark clusters with the
interaction between inert molecules.
A single quark cluster inside a quark star is assumed to be
colorless, just like each molecule in a bulk of inert gas is
electric neutral.
The interaction potential between two inert gas molecules can be
well described by the Lennard-Jones potential~\citep{LJ1924}
\begin{equation}u(r)=4U_0[(\frac{r_0}{r})^{12}-(\frac{r_0}{r})^6],\end{equation}
where $U_0$ is the depth of the potential and $r_0$ can be
considered as the range of interaction. This form of potential has
the property of short-distance repulsion and long-distance
attraction.
We assume that the interaction between the quark clusters in quark
stars can also be described by the form of Lennard-Jones potential.
\footnote{The interaction between nuclei can be described by the
$\sigma-\omega$ model~\citep{walecka1974}, which is also
characterized by the short-distance repulsion and long-distance
attraction. Recently, the nucleon-nucleon potential has been studied
by lattice QCD simulations~\citep{ishii2007}, and they also derives
a strong repulsive core at short distances.
The interaction between quark-clusters in cold quark matter could
also have long-distance attraction and short-distance repulsion if
scale and vector mesons contribute there.
We note that this short-distance repulsion is essential to reproduce
a stiff equation of state in our model.}
If the inter-cluster potential is deep enough to trap the clusters
in the potential wells, the quark matter would crystallize and form
solid quark stars.
Under such potential, we can get the equation of state for solid
quark stars, where the pressure comes from both the inter-cluster
potential and lattice vibrations.
Because the chromo-interaction is stronger than the electromagnetic
interaction that is responsible for the intermolecular forces, the
values of parameters, $U_0$ and $r_0$ used in the cold quark matter
should be different from that in the inert gas, and should be
determined in the context of quark stars.

The model of quark stars composed of Lennard-Jones matter is much
different from the conventional models (e.g., MIT bag model) in
which the ground state is of Fermi gas. In the former case the
quark-clusters are non-relativistic particles, whereas in the the
latter case quarks are relativistic particles.
Consequently, the equations of state in this two kinds of models are
different, and we found that the Lennard-Jones model has some more
stiffer equations of state, which lead to higher maximum masses for
quark stars.
Certainly, a quark star can be very low massive
($<10^{-3}M_{\odot}$) due to self color interaction.
On the other hand, we find that under some reasonable values of
parameters, a quark star could also be very massive ($>2M_{\odot}$).

This paper is arranged as follows.
The details of lattice thermodynamics are listed in Section 2.
The forms of equations of state are given in Section 3, including
the comparison with the MIT bag model, and we show the corresponding
mass-radius curves in Section 4.
We make conclusions and discussions in Section 5.

\section{Lattice thermodynamics}

For an inter-cluster potential which is deep enough to trap the
clusters in the potential wells, the quark matter would crystallize
to get lower energy and form solid quark stars.
In this section we use the results in classical solid physics to
discuss the properties of crystallized cold quark matter.

\subsection{The inter-cluster potential}

Like the inert gas, the interaction potential $u$ between two
quark-clusters as the function of their distance $r$ is described by
the Lennard-Jones potential, Eq.(1).
Let us consider a system containing $N$ clusters with the volume
$V$, then the total interaction potential is
\begin{equation}U=\frac{1}{2}\sum_i \sum_{j\neq i}u(r_{ij}),\end{equation}
and if we ignore the surface tension, we get
\begin{equation}U=\frac{N}{2}\sum_{j\neq i}u(r_{ij})=\frac{N}{2}\sum_{j\neq
i}\{4U_0[(\frac{r_0}{r_{ij}})^{12}-(\frac{r_0}{r_{ij}})^6]\}.\end{equation}
The lattice structure of cold quark matter is unknown, and we adopt
the simple-cubic structure.
The cold quark matter may have other kinds of structures, but that
will not make much differences at least quantitatively.
If the nearest distance between two quark-clusters is $R$, then the
total interaction potential of $N$ quark-clusters is
\begin{equation}U(R)=2NU_0[A_{12}(\frac{r_0}{R})^{12}-A_6(\frac{r_0}{R})^6],\end{equation}
where $A_{12}=6.2$, and $A_6=8.4$.
In the simple cubic structure, the number density of clusters $n$ is
\begin{equation}n=R^{-3},\end{equation}
so
\begin{equation}U(R)=2NU_0(A_{12}r_0^{12}n^4-A_6 r_0^6 n^2),\end{equation}
and the potential energy density is
\begin{equation}\epsilon_{\rm p}=2U_0(A_{12}r_0^{12}n^5-A_6 r_0^6 n^3),\end{equation}

\subsection{Lattice vibrations}

Consider a system of volume $V$ containing $N$ quark-clusters. Each
quark-cluster in the crystal lattice undergoes a three-dimensional
vibration about its lattice site.
Performing a normal-mode analysis in which the vibrations of the
lattice are decomposed into $3N$ independent normal modes of
vibrations, the total lattice vibration is a superposition of these
$3N$ decoupled vibrations.

For cold quark matter, the thermal vibration can be neglect compared
to the zero-point energy of phonon, so the average energy of an
individual mode of vibration with frequency $\omega_j$ is
\begin{equation}E_j=\frac{1}{2}\hbar \omega_j.\end{equation}
Using Debye approximation, at low temperature, the thermodynamics
properties of crystal lattice is mainly determined by the
long-wavelength sound waves, and the propagation of the wave can be
decomposed into one longitude mode with velocity $v_\parallel$ and
two transverse modes with velocity $v_\perp$, and the total velocity
$v$ has the relation
\begin{equation}\frac{1}{v^3}=\frac{1}{3}(\frac{1}{v_\parallel^3}+\frac{2}{v_\perp^3}),\end{equation}
and the total energy of the $3N$ vibrations is
\begin{equation}\overline{E}=\int^{\omega_m}_0 \frac{1}{2}\hbar \omega f(\omega)d\omega ,\end{equation}
where $f(\omega)d\omega$ is the number of modes in the interval from
$\omega$ to $\omega+d\omega$, and $\omega_m$ is the maximum
frequency related to the non-continuous structure of the solid, and
is determined by
\begin{equation}\omega_m=v(6\pi^2n)^{1/3}.\end{equation}
Under the condition of zero-temperature, the integration can be done
and the total energy of the crystal vibrations is
\begin{equation}\overline{E}=\frac{9V}{8}(6\pi^2)^{\frac{1}{3}}\hbar
v n^{\frac{4}{3}},\end{equation}
so the energy density of the lattice vibration is
\begin{equation}\epsilon_{\rm L}=\frac{9}{8}(6\pi^2)^{\frac{1}{3}}\hbar
v n^{\frac{4}{3}}.\end{equation}

\section{Equation of state for quark stars}

\subsection{Quark stars composed of Lennard-Jones matter}

The total energy density for cold quark matter is
\begin{eqnarray}\epsilon_{\rm q}&=&\epsilon_{\rm p}+\epsilon_{\rm L}+nm_{\rm c}c^2\nonumber\\
&=&2U_0(A_{12}r_0^{12}n^5-A_6 r_0^6
n^3)\nonumber\\&&+\frac{9}{8}(6\pi^2)^{\frac{1}{3}}\hbar v
n^{\frac{4}{3}}+nm_{\rm c}c^2,\end{eqnarray}
here $m_{\rm c}$ is the mass of each quark-cluster.
The pressure can be derived as
\begin{eqnarray}P_{\rm q}&=&n^2\frac{d(\epsilon_{\rm q}/n)}{dn}\nonumber\\
&=&4U_0(2A_{12}r_0^{12}n^5-A_6r_0^6n^3)+\frac{3}{8}(6\pi^2)^{\frac{1}{3}}\hbar
v n^{\frac{4}{3}}.\end{eqnarray}

Apart from quarks, there are electrons in quark matter.
In MIT bag model, the number of electrons per baryon $N_e/A$ is
found for different strange quarks mass $m_{\rm s}$ and coupling
constant $\alpha_{\rm s}$~\citep{FJ1984}.
In their results, when $\alpha_{\rm s}=0.3$, $N_{\rm e}/A$ is less
than $10^{-4}$; a larger $\alpha_{\rm s}$ means a smaller $N_{\rm
e}/A$ at fixed $m_{\rm s}$, because the interaction between quarks
will lead to more strange quarks and consequently less electrons.
In our model, we also consider the strong interaction between quarks
as well as between quark-clusters, and consequently the required
number of electrons per baryon to guarantee the neutrality should be
also be very small.
Although at the present stage we have not got the exactly value for
the number density of electrons, we assume that $N_{\rm e}/A$ is
less than $10^{-4}$.
We find that even at this value, the pressure of the degenerate
electrons is negligible comparing to the pressure of quarks.
Therefore, we neglect the contribution of electrons to the equation
of state.
Then the equation of state for quark stars is
\begin{equation}P=P_{\rm q},\end{equation}
\begin{equation}\rho=\epsilon_{\rm q}/c^2,\end{equation}

\subsection{Parameters}

Up to now, there are a couple of parameters in the equation of
state, and in this section we will show how to determine them.

1. The long-wavelength sound speed $v$ for lattice vibration.
For the extremely relativistic systems, the sound velocity is
$1/\sqrt{3}$, and in general it will be less than $1/\sqrt{3}$.
We find that the equation of state does not change much when $v$
goes from the velocity of light $c$ to $10^{-5}c$, so in our
calculations we set $v=c/3$.

2. The mass of quarks.
Quark stars are composed entirely of deconfined light quarks (up,
down and strange quarks), the so-called strange stars.
Because the deconfined phase transition and the chiral-restoration
phase transition might not occur simultaneously in the QCD
phase-diagram, we give each quark a constituent mass and assume it
is one-third of nuclear mass.

3. The number of quarks in one cluster $N_{\rm q}$.
Quarks are fermions and they have three flavor (up, down and
strange) degrees of freedom and three color degrees of freedom.
Pauli's exclusion principle tells us that if in the inner space
quarks are exchange-asymmetric, they are exchange-symmetric in
position space and they have a tendency to condensate in position
space.
We therefore conjecture the existence of quark-clusters in quark
matter and leave the number of quarks in one cluster, $N_{\rm q}$,
as a free parameter.
A 18-quark cluster, called quark-alpha~\citep{Michel1988}, could be
completely asymmetric in spin, flavor and color space, so in our
calculation we set $N_{\rm q}$=18, and we also set $N_{\rm q}$=3.
On the other hand, it has been conjectured that strongly interacting
matter at high densities and low temperatures might be in a
``quarkyonic'' state, which also contains three quarks in one
cluster, and is characterized by chiral symmetry and
confinement~\citep{mclerran2007,blaschke2008}.
However, in our model, the state is characterized by chiral symmetry
breaking and deconfinement.

4. The depth of the potential $U_0$.
Given the density of quark matter $\rho$ and the mass of each
individual quark, from Heisenberg's uncertainty relation we can
approximate the kinetic energy of one cluster as
\begin{equation}E_{\rm k}\sim 1\ {\rm
MeV}(\frac{\rho}{\rho_0})^{\frac{2}{3}}(\frac{N_{\rm
q}}{18})^{-\frac{5}{3}},\end{equation}
where $\rho_0$ is the nuclear matter density.
To get the quarks trapped in the potential wells to form lattice
structure, $U_0$ should be larger than the kinetic energy of quarks.
Because of the strong interaction between quarks, we adopt $U_0$=50
MeV and 100 MeV to do the calculations.

5. The range of the action $r_0$.
Since a quark star could be bound not only by gravity but also by
strong interaction due to the confinement between quarks, the number
density of quarks on a quark star surface $\rho_{\rm s}$ is
non-zero.
For a given $\rho_{\rm s}$, we can get $r_0$ at the surface where
the pressure is zero.
We choose $\rho_{\rm s}$ as 2 times of nuclear matter densities, and
get the value of $r_0$ accordingly, which is found in the range from
about 1 to 3 fm.

When $U_0$ and $r_0$ are given, the inter-cluster potential Eq.(1)
is fixed.
One should note that it describes the interaction between only two
clusters; if we consider other clusters' influences, a cluster will
always be in the minimal potential state.
When the cluster deviates from the equilibrium position, it will be
pulled back due to the stronger repulsion from one side, just as the
case of a chain of springs.

\subsection{Comparison with the MIT bag model}

In the MIT bag model, quark matter is composed of massless up and
down quarks, massive strange quarks, and few electrons.
Quarks are combined together by an extra pressure, denoted by the
bag constant $B$.
For the comparison, we apply the formulae given by
Alcock~\citep{Alcock1986} to calculate the equation of state, with
strange quark mass $m_{\rm s}=100 \rm MeV$, the strong coupling
constant $\alpha_{\rm s}=0.3$, and the bag constant $B=60 \rm MeV /
fm^{-3}$~\citep[e.g.,][]{z2000}.
The comparison of equation of state in our model and in the MIT bag
model is shown in Fig.1.
%
\begin{figure}
  \includegraphics[width=3 in]{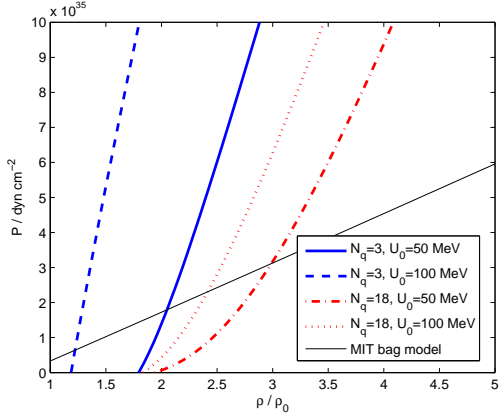}
\caption{%
The comparison of the equations of state, in the case $N_{\rm q}=3$,
including $U_0=50$ MeV (blue solid lines) and $U_0=100$ MeV (blue
dashed lines), and the corresponding case $N_{\rm q}=18$ with
$U_0=50$ MeV (red dash-dotted lines) and $U_0=100$ MeV (red dotted
lines), and that derived in the MIT bag model with the mass of
strange quark $m_{\rm s}=100$ MeV and the strong coupling constant
$\alpha_{\rm s}=0.3$ and bag constant $B=60 \rm MeV / fm^{-3}$ (thin
lines), for a given surface density $\rho_{\rm s}=2\rho_0$. Here and
in the following figures, $\rho_0$ is the nuclear saturation
density.
\label{R}}
\end{figure}
%

In our model, quarks are grouped in clusters and these clusters are
non-relativistic particles.
If the inter-cluster potential can be described as the Lennard-Jones
form, the equation of state can be very stiff, because at a small
inter-cluster distance (i.e., the number density is large enough),
there is a very strong repulsion.
Whereas in MIT bag model quarks are relativistic particles (at least
for up and down quarks).
For a relativistic system, the pressure is proportional to the
energy density, so it cannot have stiff equation of state.

\subsection{The speed of sound}

The adiabatic sound speed is defined as
\begin{equation}c_s=\sqrt{dP/d\rho}.\end{equation}
If we use the equation of state in our model, the speed of sound
will exceed the speed of light not far away from the surface of a
quark star.
It seems to contradict to the relativity that signals cannot
propagate faster than light.

The possibility of speed of sound exceeding the speed of light in
ultradense matter have been discussed previously~\citep{bludman1968}
because of using classical potential (i.e., a kind of action at a
distance). The physical reasons of apparent superluminal have also
been analyzed~\citep{caporaso1979}.
The authors argued that the adiabatic sound speed can exceed the
speed of light, yet signals propagate at speed less than $c$.

One reason is that the $P(\rho)$ relation arises from a static
calculation, ignoring the dynamics of the medium.
The notion that $c_s$ is a signal propagation speed is a carry-over
from Newtonian hydrodynamic, in which one assumes infinite
interaction speed but finite temperature, so the static and dynamic
calculations give the same result.
On the other hand, if one assumes finite interaction speed and zero
temperature, the adiabatic sound speed is not a dynamically
meaningful speed, but only a measure of the local stiffness.
Another reason is that a lattice does not have an infinte range of
allowed frequencies of vibration, but a signal should contain
components at all frequencies.
Therefore, the adiabatic sound speed is not capable of giving the
velocity of propagation of disturbs.

In our model, although we have not make it explicit that how the
particles interact with each other, we may assume that the
interaction is mediated by some particles with non-zero masses, and
the interaction does not propagate instantaneously.
We have also use the low frequency approximation to calculate the
lattice energy.
Therefore, we could conclude that in our model the signal can not
propagate faster than light.

Whether the equation of state of cold quark matter can be so stiff
that the adiabatic speed of sound is larger than $c$ could still be
an open question.
However, in our present paper we do not put the limitation on the
adiabatic sound speed, and only treat it as a measurement of the
stiffness of the equation of state.

\section{Masses and radii}

From the equations of state, we can get the mass-radius curves and
mass-central density curves (the central density only includes the
rest mass energy density), as are shown in Fig.2.

%
\begin{figure}
  \includegraphics[width=3 in]{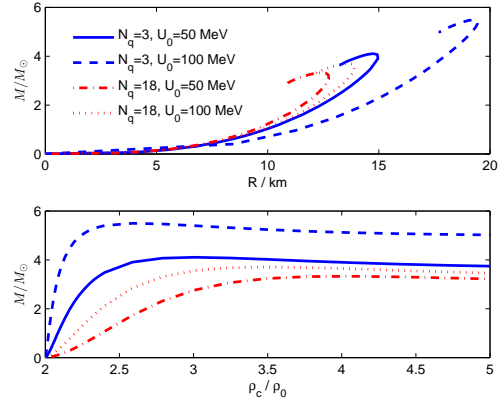}
\caption{%
The mass-radius and mass-central density (rest-mass energy density)
curves, in the case $N_{\rm q}=3$, including $U_0=50$ MeV (blue
solid lines) and $U_0=100$ MeV (blue dashed lines), and the
corresponding case $N_{\rm q}=18$ with $U_0=50$ MeV (red dash-dotted
lines) and $U_0=100$ MeV (red dotted lines), for a given surface
density $\rho_{\rm s}=2\rho_0$.
\label{R}}
\end{figure}

Because of stiffer equations of state, which we have discussed in
\S3, the maximum masses of quark stars in our model could be higher.
In Fig.2, we can see that (i) a deeper potential well $U_0$ means a
higher maximum mass; (ii) if there are more quarks in a
quark-cluster, the maximum mass of a quark star will be lower.

A stiffer equation of state leading to a higher maximum mass could
have very important astrophysical implications.
Although we could still obtain high maximum masses under MIT bag
model by choosing suitable parameters~\citep{zdunik2000}, a more
realistic equation of state in the density-dependent quark mass
model~\citep[e.g.,][]{dey1998} is very difficult to reach a high
enough maximum stellar mass, which was considered as possible
negative evidence for quark stars~\citep{cottam2002}.
Some recent observations have indicated some massive ($\sim2
M_{\odot}$) pulsars~\citep[e.g.,][]{Freire}; however, because of the
uncertain inclination of the binary systems, we are still not sure
about the real mass.
Though we have not definitely detected any pulsar whose mass is
higher than 2$M_{\odot}$ up to now, the Lennard-Jones quark star
model could be supported if massive pulsars ($>2M_\odot$) are
discovered in the future.
Moreover, a high maximum mass for quark stars might be helpful for
us to understand the mass-distribution of stellar-mass black
holes~\citep{bailyn1998}, since a compact star with a high mass
(e.g., $\sim 5M_\odot$) could still be stable in our model
presented.

\section{Conclusions and Discussions}

In cold quark matter at realistic baryon densities of compact stars
(with an average value of $\sim 2-3\rho_0$), the interaction between
quarks is so strong that they would condensate in position space to
form quark-clusters.
Like the classical solid, if the inter-cluster potential is deep
enough to trap the clusters in the potential wells, the quark matter
would crystallize and form solid quark stars.
This picture of quark stars is different from the one in which
quarks form cooper pairs and quark stars are consequently color
super-conductive.

In this paper, we argue that quarks in quark stars are grouped in
clusters and the quark-clusters form simple-cubic structure, and
apply Lennard-Jones potential to describe the interaction potential
between quark-clusters.
The parameters such as the depth of potential $U_0$ (50 MeV and 100
MeV) and the range of interaction $r_0$ (about 1 to 3 fm) are given
by the physical context of quark stars.
Under such equations of state, the masses and radii of quark stars
are derived, and we find that the mass of a quark star can be higher
than 2$M_{\odot}$.

It is surely interesting to experimentally or observationally
distinguish between our solid quark star model and other models for
quark stars, e.g., the color super-conductivity state.
Starquakes could naturally occur in solid quark stars and the
observations of pulsar glitches and SGR giant flares could
qualitatively be reproduced when the solid matter
breaks~\citep{zhou2004,xu06}, moreover the post-glitch recovers in
the solid quark star model and the color super-conductivity model
would be different.
Additionally, because the solid quark star model depends on quark
clustering, the interaction behaviors between quarks could be tested
in sQGP (strongly coupled quark-gluon plasma, see~Shuryak 2009) by
the LHC and/or FAIR experiments.

The research of compact stars involves two kinds of challenges:
particle physics and many-body physics.
Nevertheless, if we know about the properties of compact stars from
observations, we can get information of the elementary physics.
Take the model we discussed in this paper as an example.
If we get the masses and radii of some pulsars from accurate enough
observations, we can put limits on the parameters such as potential
well depth $U_0$, interaction range $r_0$, and the number of quarks
that condensate in position space to form a cluster, which could
help us to explore the strong interaction between quarks.
Although the state of cold quark matter at a few nuclear densities
is still an unsolved problem in the low-energy QCD, it would be
hopeful for us to use pulsars as idea laboratories to study the
nature of strong interaction.

In general, stars are equilibrium bodies with pressure against
gravity.
The thermal and radiation pressure dominates in main sequent stars,
while degenerate pressure of Fermions, originated from Pauli's
principle, dominates in Fermion stars (e.g., while dwarfs).
For solid quark stars in the models presented in this paper, the
pressure is related to the increase of both potential and lattice
vibration energies as the stellar quark matter contracts. The
degenerate pressure might be negligible there.

\section*{Acknowledgments}
\thanks{%
We would like to acknowledge useful discussions with Prof. Rachid
Ouyed of University of Calgary and members at our pulsar group of
PKU, and to thank an anonymous referee for valuable comments and
suggestions.
This work is supported by NSFC (10778611), the National Basic
Research Program of China (grant 2009CB824800) and by LCWR
(LHXZ200602).}

\end{document}